\documentclass{article}
\usepackage{hyperref}
\usepackage{authblk}
\usepackage[utf8]{inputenc}
\usepackage{booktabs}
\usepackage{amsmath}
\usepackage{cleveref}
\usepackage{paralist}
\usepackage{enumitem}
\usepackage{csquotes}
\usepackage{xcolor}
\usepackage[numbers,sort]{natbib}
\usepackage{pf2}

\numberwithin{equation}{section}
\hypersetup{
    colorlinks,
    linkcolor={red!50!blue},
    citecolor={red!50!blue},
    urlcolor={red!50!blue}
}

\def\tlathin{\mskip.5\thinmuskip}
\def\indentspace#1{\def\tla@indentspace{#1}}
\indentspace{1em}

\newcommand{\deq}{\mathrel{\stackrel{\scriptscriptstyle\Delta}{=}}}
\def\A{\forall\,}

\newcommand{\tupL}{\ensuremath{\langle\tlathin}}
\newcommand{\tupR}{\ensuremath{\tlathin\rangle}}

\def\IF#1{\mbox{{\sc if\, }$#1$ }%
             \begin{array}[t]{@{}l@{}l@{}}}
\def\THEN{\mbox{\sc \,then\, }&\begin{array}[t]{@{}l@{}}}
\def\ELSE{\end{array}\\\mbox{\sc \,else\, }&\begin{array}[t]{@{}l@{}}}
\def\FI{\end{array}\end{array}}

\newcommand{\LET}{\begin{array}[t]{@{}l@{\;\;}l@{}} \mbox{\sc let}%
                              &\begin{array}[t]{@{}l@{}}}
\newcommand{\IN}{\end{array}\\\mbox{\sc in}&\begin{array}[t]{@{}l@{}}}
\newcommand{\NI}{\end{array}\end{array}}

\newcommand{\EXCEPT}{\mbox{\sc\ except }}

\newcommand{\FALSE}{\mbox{\sc false}}

\newcommand{\Deq}{\ensuremath{\;\;\deq\;\;}}

\newcommand{\I}{\mathbf{I}}
\newcommand{\bft}{\mathbf{t}}
\newcommand{\bfs}{\mathbf{s}}
\newcommand{\bfV}{\mathbf{V}}

\newcommand{\ra}[1]{\renewcommand{\arraystretch}{#1}}
\ra{1.2}

\newcommand{\allnotes}[1]{}
\renewcommand{\allnotes}[1]{\textit{#1}}

\title{A Decidable Case of Query Determinacy: Project-Select Views}
\author[1]{Wen Zhang}
\author[2]{Aurojit Panda}
\author[3]{Mooly Sagiv}
\author[1,4]{Scott Shenker}
\affil[1]{UC Berkeley}
\affil[2]{NYU}
\affil[3]{Tel Aviv University}
\affil[4]{ICSI}
\date{}

\newtheorem{theorem}{Theorem}
\newtheorem{corollary}[theorem]{Corollary}

\begin{document}

\maketitle

\begin{abstract}
    Query determinacy is decidable for project-select views and a project-select-join query with no self joins---as long as the selection predicates are in a first-order theory for which satisfiability is decidable.
\end{abstract}

\section{Summary}
To enforce a view-based access-control policy, the Blockaid system checks a property on SQL views and queries that generalizes \emph{query determinacy} under set semantics~\cite[\S~4.2]{DBLP:conf/osdi/ZhangSCPSS22}.
A set~$\bfV$ of queries (which we will call ``views'') \emph{determines} a query~$Q$ if and only if $\bfV(\I)=\bfV(\I')$ implies $Q(\I)=Q(\I')$~\cite{DBLP:journals/tods/NashSV10}.%
\footnote{We may think of $\bfV(I)$ as a database instance that, for each $V\in \bfV$, maps the relation name~$V$ to the relation~$V(\I)$.  Then the formula $\bfV(\I)=\bfV(\I')$ means $\A V\in\bfV : V(\I)=V(\I')$.}

Checking query determinacy is a hard problem---it is undecidable even for CQ (conjunctive query) views and CQ queries~\cite{DBLP:conf/lics/GogaczM15,DBLP:conf/pods/GogaczM16}.
It \emph{is} shown to be decidable in simpler scenarios---for example,
\begin{inparaenum}[(1)]
    \item for MCQ (monadic-conjunctive-query) views and CQ queries~\cite[Theorem~5.16]{DBLP:journals/tods/NashSV10}, and
    \item for a single path-query view and a CQ query~\cite[Theorem~5.20]{DBLP:journals/tods/NashSV10}.
\end{inparaenum}
But these decidability results are too limited to be applied to view-based access-control for practical web applications.

Here, we discuss another case where query determinacy is decidable: for select-project views, and a select-project-join query with no self joins---as long as the selection predicates are not too complex.
To be clear, this result is still quite limited, for the simple reason that real-world views often have joins.
But it is a step forward in our search for a larger class of views and queries, encompassing more real-world use cases, for which query determinacy is decidable.

\section{Setup}
We consider all queries under set semantics, and 
we assume that all select-project-join queries are put into normal form~\cite[Proposition~4.4.2]{DBLP:books/aw/AbiteboulHV95}.

Fix a database schema consisting of relation names $R_1, R_2, \ldots, R_m$, and let $\bfV$ be a set of views.
Since we're dealing only with project-select views, we partition $\bfV$ into $\bfV_1, \bfV_2, \ldots, \bfV_m$,
where each $\bfV_i$ consists of some number~$n_i$ of queries that refer only to the relation~$R_i$.
In other words, we denote:
\begin{align}
    \bfV &\Deq \bfV_1 \cup \bfV_2 \cup \cdots \bfV_m, \nonumber \\
    \bfV_i &\Deq \{ V_{i,j} : 1 \leq j \leq n_i \}, & (1\leq i\leq m) \nonumber \\
    V_{i,j} &\Deq \pi_{U_{i,j}} \sigma_{\theta_{i,j}} R_i. & (1\leq i\leq m, 1\leq j\leq n_i)\label{eqn:vij}
\end{align}
And let query~$Q$ be a project-select-join query with no self joins:
\begin{equation}\label{eqn:q}
    Q \Deq \pi_U \sigma_{\theta} (R_1\times R_2\times \cdots\times R_m).
\end{equation}

\begin{table}
    \caption{Database notations.}\label{tab:db-notations}
    \begin{center}
        \begin{tabular}{rl}
            \toprule
            $m$ & Number of relations in the database schema \\
            $R_i$ & Name of relation \\
            $\I$, $\I'$ & Database instance (a mapping from relation names to relations) \\
            $V_{i,j}$, $Q$ & Query \\
            $\bfV$, $\bfV_i$ & Set of queries \\
            $U$, $U_{i,j}$ & Set of column names \\
            $\theta$, $\theta_{i,j}$ & Predicate (used in selections) \\
            $t$ & Database tuple (a mapping from column names to values) \\
            $t_i$, $t_{i,j}$ & Tuple in relation~$R_i$ \\
            $t[U]$ & Sub-tuple $\{ A \mapsto t(A) : A \in U \}$ \\
            \bottomrule
        \end{tabular}
    \end{center}
\end{table}

\begin{table}
    \caption{Other mathematical notations.}\label{tab:other-notations}
    \begin{center}
        \begin{tabular}{rl}
            \toprule
            $\bft$ & Sequence of tuples $\tupL t_1, \ldots, t_{|\bft|} \tupR$ \\
            $\bft_{m..n}$ & Sub-sequence of tuples $\tupL t_m, t_{m+1}, \ldots, t_n \tupR$ \\
            $\bft \circ \bfs$ & Concatenation of the sequence $\mathbf{t}$ and the sequence $\bfs$ \\
            $\bft \EXCEPT i \mapsto s$ & Sequence of tuples $\tupL t_1, \ldots, t_{i-1}, s, t_{i+1}, \ldots, t_{|\bft|} \tupR$ \\
            \bottomrule
        \end{tabular}
    \end{center}
\end{table}

A summary of notations is found in~\Cref{tab:db-notations,tab:other-notations}.
For visual clarity, we will use lists bulleted by ``$\land$'' to denote the conjunction of a number of formulas~\cite{DBLP:journals/fac/Lamport94}.

\section{Reducing determinacy to a logical formula}
In this setting, checking whether $\bfV$ determines $Q$ can be reduced to checking the satisfiability of a logical formula.

\begin{theorem}\label{thm:determine}
    The set~$\bfV$ of views determines query~$Q$ iff for every $1\leq i\leq m$:
    \begin{equation}\label{eqn:determine}
        \A \bft : \theta(\bft) \Rightarrow
            \bigvee_{j=1}^{n_i} \left(
                \theta_{i,j}(t_i) \land \A t_i' : \Phi_{i,j}(t_i,t_i') \Rightarrow \Psi_{i,j}(\bft,t_i')
            \right) \tag{$\star$}
    \end{equation}
    where the sub-formulas $\Phi$ and $\Psi$ are defined as:
    \begin{align*}
        \Phi_{i,j}(t_i,t_i') &\Deq
        \begin{conj}
            \theta_{i,j}(t_i') \\
            t_i'[U_{i,j}]=t_i[U_{i,j}],
        \end{conj} \\
        \Psi_{i,j}(\bft,t_i') &\Deq
        \LET \bfs \Deq \bft \EXCEPT i \mapsto t_i'
        \IN
            \begin{conj}
                \theta(\mathbf{s})\\
                \mathbf{s}[U]=\bft[U].
            \end{conj}
        \NI
    \end{align*}
    Furthermore, finite and unrestricted determinacy coincide in this setting.
\end{theorem}

We present its proof in \Cref{sec:proof}.

As a result of this theorem, for project-select views and a project-select-join query without self joins, we can check query determinacy by checking the validity of~(\ref{eqn:determine}) for every $1\leq i\leq m$.
That is, query determinacy is decidable as long as the validity of~(\ref{eqn:determine}) is decidable.

Observe that (\ref{eqn:determine}) is constructed from selection predicates~$\theta$ and $\theta_{i,j}$, equality atoms, and propositional connectives.

\begin{corollary}
    For project-select views and a project-select-join query without self joins,
    if all selection predicates in the views and the query are in a first-order theory for which validity is decidable,
    then query determinacy is decidable.
\end{corollary}

For example, if the selection predicates consist only of equalities over variables and propositional connectives~\cite[\S~9]{DBLP:books/daglib/0019162}, then query determinacy is decidable.
A straightforward way to check determinacy is by encoding~(\ref{eqn:determine}) into an SMT formula and calling an SMT solver.
What's more, if the selection predicates are quantifier-free (which they typically are in practice),
then the negation of (\ref{eqn:determine}) is purely existentially quantified, so we can check determinacy simply by checking the satisfiability of a formula that is quantifier-free.

\bibliography{ref}{}

\begin{thebibliography}{9}
\providecommand{\natexlab}[1]{#1}
\providecommand{\url}[1]{\texttt{#1}}
\expandafter\ifx\csname urlstyle\endcsname\relax
  \providecommand{\doi}[1]{doi: #1}\else
  \providecommand{\doi}{doi: \begingroup \urlstyle{rm}\Url}\fi

\bibitem[Abiteboul et~al.(1995)Abiteboul, Hull, and
  Vianu]{DBLP:books/aw/AbiteboulHV95}
Serge Abiteboul, Richard Hull, and Victor Vianu.
\newblock \emph{Foundations of Databases}.
\newblock Addison-Wesley, 1995.
\newblock ISBN 0-201-53771-0.
\newblock URL \url{http://webdam.inria.fr/Alice/}.

\bibitem[Bradley and Manna(2007)]{DBLP:books/daglib/0019162}
Aaron~R. Bradley and Zohar Manna.
\newblock \emph{The calculus of computation - decision procedures with
  applications to verification}.
\newblock Springer, 2007.
\newblock \doi{10.1007/978-3-540-74113-8}.
\newblock URL \url{https://doi.org/10.1007/978-3-540-74113-8}.

\bibitem[Gogacz and Marcinkowski(2015)]{DBLP:conf/lics/GogaczM15}
Tomasz Gogacz and Jerzy Marcinkowski.
\newblock The hunt for a red spider: Conjunctive query determinacy is
  undecidable.
\newblock In \emph{30th Annual {ACM/IEEE} Symposium on Logic in Computer
  Science, {LICS} 2015, Kyoto, Japan, July 6-10, 2015}, pages 281--292. {IEEE}
  Computer Society, 2015.
\newblock \doi{10.1109/LICS.2015.35}.
\newblock URL \url{https://doi.org/10.1109/LICS.2015.35}.

\bibitem[Gogacz and Marcinkowski(2016)]{DBLP:conf/pods/GogaczM16}
Tomasz Gogacz and Jerzy Marcinkowski.
\newblock Red spider meets a rainworm: Conjunctive query finite determinacy is
  undecidable.
\newblock In Tova Milo and Wang{-}Chiew Tan, editors, \emph{Proceedings of the
  35th {ACM} {SIGMOD-SIGACT-SIGAI} Symposium on Principles of Database Systems,
  {PODS} 2016, San Francisco, CA, USA, June 26 - July 01, 2016}, pages
  121--134. {ACM}, 2016.
\newblock \doi{10.1145/2902251.2902288}.
\newblock URL \url{https://doi.org/10.1145/2902251.2902288}.

\bibitem[Lamport(1994)]{DBLP:journals/fac/Lamport94}
Leslie Lamport.
\newblock How to write a long formula (short communication).
\newblock \emph{Formal Aspects Comput.}, 6\penalty0 (5):\penalty0 580--584,
  1994.
\newblock \doi{10.1007/BF01211870}.
\newblock URL \url{https://doi.org/10.1007/BF01211870}.

\bibitem[Lamport(2006)]{DBLP:journals/dc/Lamport06a}
Leslie Lamport.
\newblock Lower bounds for asynchronous consensus.
\newblock \emph{Distributed Comput.}, 19\penalty0 (2):\penalty0 104--125, 2006.
\newblock \doi{10.1007/S00446-006-0155-X}.
\newblock URL \url{https://doi.org/10.1007/s00446-006-0155-x}.

\bibitem[Lamport(2012)]{lamport_how_2012}
Leslie Lamport.
\newblock How to write a 21st century proof.
\newblock \emph{Journal of Fixed Point Theory and Applications}, 11\penalty0
  (1):\penalty0 43--63, March 2012.
\newblock ISSN 1661-7746.
\newblock \doi{10.1007/s11784-012-0071-6}.
\newblock URL \url{https://doi.org/10.1007/s11784-012-0071-6}.

\bibitem[Nash et~al.(2010)Nash, Segoufin, and
  Vianu]{DBLP:journals/tods/NashSV10}
Alan Nash, Luc Segoufin, and Victor Vianu.
\newblock Views and queries: Determinacy and rewriting.
\newblock \emph{{ACM} Trans. Database Syst.}, 35\penalty0 (3):\penalty0
  21:1--21:41, 2010.
\newblock \doi{10.1145/1806907.1806913}.
\newblock URL \url{https://doi.org/10.1145/1806907.1806913}.

\bibitem[Zhang et~al.(2022)Zhang, Sheng, Chang, Panda, Sagiv, and
  Shenker]{DBLP:conf/osdi/ZhangSCPSS22}
Wen Zhang, Eric Sheng, Michael~Alan Chang, Aurojit Panda, Mooly Sagiv, and
  Scott Shenker.
\newblock Blockaid: Data access policy enforcement for web applications.
\newblock In Marcos~K. Aguilera and Hakim Weatherspoon, editors, \emph{16th
  {USENIX} Symposium on Operating Systems Design and Implementation, {OSDI}
  2022, Carlsbad, CA, USA, July 11-13, 2022}, pages 701--718. {USENIX}
  Association, 2022.
\newblock URL
  \url{https://www.usenix.org/conference/osdi22/presentation/zhang}.

\end{thebibliography}
\bibliographystyle{plainnat}

\appendix
\section{Proof of Theorem~\ref*{thm:determine}}\label{sec:proof}
\Cref{thm:determine} is somewhat tedious to prove.
In an attempt to avoid mistakes and make the proof easier to check,
we write the proof in the \emph{hierarchically structured style}~\cite{lamport_how_2012}.
Explaining this style, \citet[Appendix~A]{DBLP:journals/dc/Lamport06a} writes:
\begin{displayquote}
A structured proof consists of a sequence of statements and their proofs;
each of those proofs is either a structured proof or an ordinary paragraph-style proof.
The $j$\textsuperscript{th} step in the current level~$i$ proof is numbered $\langle i\rangle j$.
\textelp{}
We recommend reading the proofs hierarchically, from the top level down. To read the proof of a long level~$i$ step, first read the level $i+1$ statements that form its proof, together with the proof of the final Q.E.D.~step (which is usually a short paragraph).
\end{displayquote}

\noindent{\bf Proof of \Cref{thm:determine}}:

\begin{proof}
    \step{<1>1}{
        \assume{(\ref{eqn:determine}) holds for every $1\leq i\leq m$.}
        \prove{$\bfV$ determines $Q$.}
    }
    \begin{proof}
        \step{<2>1}{
            \sassume{
                \begin{enumerate}
                    \item \pfnew\ $(\I, \I')$
                    \item $\bfV(\I)=\bfV(\I')$\label{item:v-i}
                    \item \pfnew\ $s\in Q(\I)$\label{item:new-s}
                \end{enumerate}
            }
            \prove{$s\in Q(\I')$}
        }
        \begin{proof}
            \pf\ By definition of query determinacy, and by symmetry between $\I$ and $\I'$.
        \end{proof}
        \step{<2>2}{
            Choose~$\bft$ such that:
            \begin{enumerate}
                \item $\theta(\bft)$ holds,\label{item:choose-bft-theta}
                \item $\bft[U]=s$,\label{item:choose-bft-U}
                \item $t_i\in \I(R_i)$ for all $1\leq i\leq m$.\label{item:choose-bft-R}
            \end{enumerate}
        }
        \begin{proof}
            \pf\ Such $\bft$ exists by \stepref{<2>1}.\ref*{item:new-s}, and by definition of~$Q$ (\ref{eqn:q}).
        \end{proof}
        \step{<2>3}{
            For every $0 \leq\ell \leq m$, there exists a sequence $\bft^{\ell}$ of $\ell$~tuples, such that:
            \begin{enumerate}[label={C\arabic*.}, ref={C\arabic*}]
                \item $t^{\ell}_i \in \I'(R_i)$ for every $1\leq i\leq \ell$,\label{c1}
                \item $\theta(\bft^{\ell} \circ \bft_{\ell+1 .. m})$ holds,\label{c2}
                \item $(\bft^{\ell} \circ \bft_{\ell+1 .. m})[U]=s$.\label{c3}
            \end{enumerate}
        }
        \begin{proof}
            We will proceed by induction on~$\ell$.
            \step{<3>1}{There exists $\bft^0$ which satisfies \ref{c1}--\ref{c3} for $\ell=0$.}
            \begin{proof}
                \pf\ Take $\bft^0$ to be the empty sequence.
                    \ref{c1} holds vacuously; \ref{c2} follows from \stepref{<2>2}.\ref*{item:choose-bft-theta};
                    and \ref{c3} follows from \stepref{<2>2}.\ref*{item:choose-bft-U}.
            \end{proof}
            \step{<3>2}{
                \assume{\pfnew\ $\bft^{r-1}$ ($0<r\leq m$), $\bft^{r-1}$ satisfies \ref{c1}--\ref{c3} for $\ell=r-1$.}
                \prove{There exists $\bft^r$ which satisfies \ref{c1}--\ref{c3} for $\ell=r$.}
            }
            \begin{proof}
                \step{<4>1}{
                    There exists $1\leq j\leq n_r$ such that:
                    \begin{align}
                        \land\ & \theta_{r,j}(t_r)\label{eqn:t1} \\
                        \land\ & \A t_r' : \Phi_{r,j}(t_r,t_r') \Rightarrow
                            \Psi_{r,j}\left((\bft^{r-1} \circ \bft_{r..m}),t_r'\right).\label{eqn:t2}
                    \end{align}
                }
                \begin{proof}
                    \pf\ By taking (\ref{eqn:determine}) with $i=r$,
                    instantiating the outermost universal quantifier using
                    $\bft^{r-1} \circ \bft_{r..m}$,
                    and applying \ref{c2} with $\ell=r-1$.
                \end{proof}
                \step{<4>2}{$t_r[U_{r,j}] \in V_{r,j}(\I)$}
                \begin{proof}
                    \pf\ $t_r\in \I(R_r)$ by \stepref{<2>2}.\ref*{item:choose-bft-R}; by (\ref{eqn:t1}); and by definition of $V_{r,j}$.
                \end{proof}
                \step{<4>3}{$t_r[U_{r,j}] \in V_{r,j}(\I')$}
                \begin{proof}
                    \pf\ By \stepref{<4>2}, noting that $V_{r,j}(\I)=V_{r,j}(\I')$ by \stepref{<2>1}.\ref*{item:v-i}.
                \end{proof}
                \step{<4>4}{
                    There exists tuple $t_r'\in \I'(R_r)$ such that $\Phi_{r,j}(t_r,t_r')$ holds.
                }
                \begin{proof}
                    \pf\ By \stepref{<4>3}, plugging in the definition of $V_{r,j}$ (\ref{eqn:vij}), there exists $t_r'\in \I'(R_r)$ such that:
                        \[
                            \begin{conj}
                                \theta_{r,j}(t_r')\\
                                t_r'[U_{r,j}] = t_r[U_{r,j}]
                            \end{conj}
                        \]
                        which implies $\Phi_{r,j}(t_r,t_r')$.
                \end{proof}
                \step{<4>5}{
                    $\Psi_{r,j}\left((\bft^{r-1} \circ \bft_{r..m}),t_r'\right)$ holds.
                }
                \begin{proof}
                    \pf\ Instantiate universal quantifier in (\ref{eqn:t2}) with $t_r'$ from \stepref{<4>4}.
                \end{proof}
                \define{$\bft^r \Deq \bft^{r-1} \circ \tupL t_r' \tupR$}
                \step{<4>6}{$\bft^r$ satisfies \ref{c1} for $\ell=r$.}
                \begin{proof}
                    \pf\ From \stepref{<3>2}, $\bft^{r-1}$ satisfies \ref{c1} for $\ell=r-1$;
                        from \stepref{<4>4}, $t_r'\in \I'(R_r)$.
                \end{proof}
                \step{<4>7}{$\bft^r$ satisfies \ref{c2} for $\ell=r$.}
                \begin{proof}
                    \pf\ By \stepref{<4>5}, plugging in definition for~$\Psi$.
                \end{proof}
                \step{<4>8}{$\bft^r$ satisfies \ref{c3} for $\ell=r$.}
                \begin{proof}
                    \pf\ By \stepref{<4>5}, plugging in definition for~$\Psi$; by \stepref{<2>2}.\ref*{item:choose-bft-U}, $\bft[U]=s$.
                \end{proof}
                \qedstep
                \begin{proof}
                    \pf\ By \stepref{<4>6}, \stepref{<4>7}, and \stepref{<4>8}.
                \end{proof}
            \end{proof}
            \qedstep
            \begin{proof}
                \pf\ By induction on~$\ell$, with \stepref{<3>1} as the base case and \stepref{<3>2} as the inductive step.
            \end{proof}
        \end{proof}
        \qedstep
        \begin{proof}
            \pf\ By \stepref{<2>3}, taking $\ell=m$; and by definition of~$Q$ (\ref{eqn:q}).
        \end{proof}
    \end{proof}
    \step{<1>2}{
        \assume{For some $1\leq k\leq m$, (\ref{eqn:determine}) does not hold for $i=k$.}
        \prove{There exist finite $\I, \I'$ such that $\bfV(\I)=\bfV(\I')$ but $Q(\I)\neq Q(\I')$.}
    }
    \begin{proof}
        \step{<2>1}{
            Choose $\bft$ such that $\theta(\bft)$ holds, but for every $1\leq j\leq n_k$:
            \begin{enumerate}[label={N\arabic*.}, ref={N\arabic*}]
                \item $\theta_{k,j}(t_k)$ does not hold, or
                \item There exists $t'_{k,j}$ such that $\Phi_{k,j}(t_k,t'_{k,j})$ but $\lnot \Psi_{k,j}(\bft,t'_{k,j})$.\label{right2}
            \end{enumerate}
        }
        \begin{proof}
            \pf\ Such $\bft$ exists by the negation of (\ref{eqn:determine}).
        \end{proof}
        \define{Database instances~$\I,\I'$:
            \begin{align*}
                \I(R_i) &\Deq 
                    \IF{i=k} \THEN{\{ t_{k,j}' : \theta_{k,j}(t_k), 1\leq j\leq n_k \}}
                    \ELSE{\{t_i\}}, \FI
                        & (\text{as in~\ref{right2}})
                \\
                \I'(R_i) &\Deq
                    \IF{i=k} \THEN{\I(R_k) \cup \{t_k\}}
                    \ELSE{\I(R_i)}. \FI
            \end{align*}
        }
        \step{<2>2}{
            \assume{\pfnew\ $V_{i,j}\in\bfV$}
            \prove{$V_{i,j}(\I)=V_{i,j}(\I')$}
        }
        \begin{proof}
            \step{<3>1}{\case{$i\neq k$}}
            \begin{proof}
                \pf\ $V_{i,j}$ refers only to relation~$R_i$ by assumption, and $\I(R_i)=\I'(R_i)$ by the construction of~$\I'$.
            \end{proof}
            \step{<3>2}{\case{$i=k$}}
            \begin{proof}
                Since $V_{k,j}$ refers only to relation name~$R_k$, we will treat $V_{k,j}$ as a function on the relation~$R_k$---i.e., $V_{k,j}(\I)=V_{k,j}(\I(R_k))$.
                \step{<4>1}{$V_{k,j}(\I(R_k)) \subseteq V_{k,j}(\I'(R_k))$}
                \begin{proof}
                    \pf\ Because $\I(R_k) \subseteq \I'(R_k)$ by definition,
                        and by monotonicity of project-select queries.
                \end{proof}
                \step{<4>2}{$V_{k,j}(\I'(R_k)) \subseteq V_{k,j}(\I(R_k))$}
                \begin{proof}
                    \step{<5>1}{\suffices{$V_{k,j}(\{ t_k \}) \subseteq V_{k,j}(\I(R_k))$}}
                    \begin{proof}
                        \pf\ Because $\I'(R_k)=\I(R_k)\cup \{t_k\}$ by construction.
                    \end{proof}
                    \step{<5>2}{\case{$\theta_{k,j}(t_k)$ holds.}}
                    \begin{proof}
                        By the definition of~$\I$,
                        there exists $t'_{k,j}\in \I(R_k)$ such that $\Phi_{k,j}(t_k,t'_{k,j})$ holds.
                        Expanding the definition of~$\Phi$, we have:
                        \[
                            \begin{conj}
                                \theta_{k,j}(t'_{k,j}) \\
                                t'_{k,j}[U_{k,j}]=t_k[U_{k,j}]
                            \end{conj}
                        \]
                        which implies $V_{k,j}(\{t_k\})=V_{k,j}(\{t'_{k,j}\}) \subseteq V_{k,j}(\I(R_k))$.
                    \end{proof}
                    \step{<5>3}{\case{$\theta_{k,j}(t_k)$ does not hold.}}
                    \begin{proof}
                        \pf\ $V_{k,j}(\{ t_k \})=\emptyset \subseteq V_{k,j}(\I(R_k))$ by definition of $V_{k,j}$ (\ref{eqn:vij}).
                    \end{proof}
                    \qedstep
                    \begin{proof}
                        \pf\ \stepref{<5>2} and \stepref{<5>3} cover all the cases.
                    \end{proof}
                \end{proof}
                \qedstep
                \begin{proof}
                    \pf\ By \stepref{<4>1} and \stepref{<4>2}.
                \end{proof}
            \end{proof}
            \qedstep
            \begin{proof}
                \pf\ \stepref{<3>1} and \stepref{<3>2} cover all the cases.
            \end{proof}
        \end{proof}
        \step{<2>3}{$Q(\I)\neq Q(\I')$.}
        \begin{proof}
            \step{<3>1}{$\bft[U] \in Q(\I')$}
            \begin{proof}
                \pf\ By construction of~$\I'$, we have $t_i\in \I'(R_i)$ for all $1\leq i\leq m$.
                    By assumption, $\theta(\bft)$ holds.
            \end{proof}
            \step{<3>2}{$\bft[U] \not\in Q(\I)$}
            \begin{proof}
                \step{<4>1}{
                    \sassume{
                        \begin{enumerate}
                            \item \pfnew\ $\bfs\in \I(R_1)\times \cdots\times \I(R_m)$
                            \item $\theta(\bfs) \land \left( \bfs[U]=\bft[U] \right)$\label{item:s-produces-t}
                        \end{enumerate}
                    }
                    \prove{\FALSE}
                }
                \begin{proof}
                    \pf\ By definition of~$Q$ (\ref{eqn:q}).
                \end{proof}
                \step{<4>2}{
                    Choose $1\leq j\leq n_k$ and tuple $t'_{k,j}$ such that:
                    \begin{enumerate}
                        \item $\bfs = \bft \EXCEPT k \mapsto t'_{k,j}$,
                        \item $\lnot \Psi_{k,j}(\bft,t'_{k,j})$.\label{item:not-psi}
                    \end{enumerate}
                }
                \begin{proof}
                    \pf\ Such $j$ and $t'_{k,j}$ exist by the construction of~$\I$.
                \end{proof}
                \qedstep
                \begin{proof}
                    \pf\ \stepref{<4>2}.\ref*{item:not-psi} contradicts \stepref{<4>1}.\ref*{item:s-produces-t}.
                \end{proof}
            \end{proof}
            \qedstep
            \begin{proof}
                \pf\ By \stepref{<3>1} and \stepref{<3>2}.
            \end{proof}
        \end{proof}
        \qedstep
        \begin{proof}
            \pf\ By \stepref{<2>2} and~\stepref{<2>3}, noting that $\I$ and $\I'$ are finite.
        \end{proof}
    \end{proof}
    \step{<1>3}{Finite and unrestricted determinacy coincide.}
    \begin{proof}
        \pf\ If unrestricted determinacy does not hold,
            by \stepref{<1>2} there exists a finite counterexample,
            and so finite determinacy does not hold either.
    \end{proof}
    \qedstep
    \begin{proof}
        \pf\ By \stepref{<1>1}, \stepref{<1>2}, and~\stepref{<1>3}.
    \end{proof}
\end{proof}

\end{document}